\documentclass[twocolumn,amssymb, nofootinbib, aps, pre]{revtex4-2}

\usepackage{amssymb,amsthm,amsmath,float}

\usepackage[total={6.5in,8.75in}, top=1.2in, left=0.9in, includefoot]{geometry}
\usepackage{graphicx}
\usepackage[pdftex,bookmarks, colorlinks, citecolor =blue, breaklinks]{hyperref}
\usepackage{url}

\newcommand{\Eq}[1]{(\ref{eq:#1})}

\newcommand{\Sec}[1]{\S \ref{sec:#1}}
\newcommand{\Fig}[1]{Fig.~\ref{fig:#1}}

\newcommand{\InsertFig}[4]
{\begin{figure}[h!t]
       \centerline{
         \includegraphics[width=#4]{./#1}
       }
       \caption{{\footnotesize  #2}
       \label{fig:#3}}
\end{figure}}

\newcommand{\InsertFigTwo}[5] {
\begin{figure*}[h!t]
       \centerline{
         \includegraphics[width=#5]{./#1}
         \hskip 0.0in
         \includegraphics[width=#5]{./#2}
       }
       \caption{{\footnotesize  #3}
       \label{fig:#4}}
\end{figure*}
}

\newcommand{\InsertFigFour}[7] {
\begin{figure*}[h!t]
       \centerline{
\renewcommand{\arraystretch}{0.01}
         \begin{tabular}{cc}
         \includegraphics[width=#7]{./#1}&  \includegraphics[width=#7]{./#2} \\
        \includegraphics[width=#7]{./#3}  &  \includegraphics[width=#7]{./#4}
        \end{tabular}
       }
       \caption{{\footnotesize  #5}
       \label{fig:#6}}
\end{figure*}
}

\newcommand{\bN}{{\mathbb{ N}}}

\newcommand{\bR}{{\mathbb{ R}}}

\newcommand{\bT}{{\mathbb{ T}}}
\newcommand{\bZ}{{\mathbb{ Z}}}

\newcommand{\cO}{{\cal O}}

\newcommand{\WB} {\mathit{WB}}

\newcommand{\beq}[1]{\begin{equation}\label{eq:#1}}
\newcommand{\eeq}{\end{equation}}

\newenvironment{se}[1]{\equation\label{eq:#1}\aligned}{\endaligned\endequation}
\newcommand{\bsplit}[1]{\begin{se}{#1}}
\newcommand{\esplit}{\end{se}}

\newenvironment{example}[1][]
  {
	\setlength \leftmargini {1.0em}		
	\setlength \topsep {0.5em}			
	\begin{quote}
	{\it Example#1} }
	{\end{quote}
  }
\newcommand{\bexam}[1][:]{\begin{example}[#1]}
\newcommand{\eexam}{\end{example}}

\begin{document}
\title{Computing Lyapunov Exponents using Weighted Birkhoff Averages}
\author{E. Sander}
\email{esander@gmu.edu}
\affiliation{Department of Mathematical Sciences, George Mason University, Fairfax, VA 22030, USA}
\thanks{ ES was supported in part by Simons Grant 636383.}
\author{J.D.~Meiss}
\email{jdm@colorado.edu}
\affiliation{Department of Applied Mathematics, University of Colorado,  Boulder, CO 80309-0526, USA}
\thanks{JDM was supported in part by Simons Grant 601972.}

\date{\today}

\begin{abstract}
The Lyapunov exponents of a dynamical system measure the average rate of exponential 
stretching along an orbit. Positive exponents are often taken as a defining characteristic of chaotic dynamics.
However, the standard orthogonalization-based method for computing Lyapunov exponents converges slowly---if at all.
Many alternatively techniques have been developed to distinguish between regular and chaotic orbits,
though most do not compute the exponents.
We compute the Lyapunov spectrum in three ways: the standard method, the weighted Birkhoff average (WBA),
and the ``mean exponential growth rate for nearby orbits'' (MEGNO). 
The latter two improve convergence for nonchaotic orbits, but the WBA is fastest. 
However, for chaotic orbits the three methods convergence at similar, slow rates. 
Though the original MEGNO method does not compute Lyapunov exponents, we show how
to reformulate it as a weighted average that does.
\end{abstract}



\maketitle

\section{Introduction}

Lyapunov exponents are a fundamental gauge of chaotic behavior in dynamical systems.
They measure the growth rate of the distance between a pair of (infinitesimally) close orbits,
and a positive exponent is often taken as a primary indicator for ``sensitive dependence
on initial conditions,'' one of the principal requirements for chaos.
Formally, given a differentiable map $f: M \to M$ on an $n$-dimensional phase space $M$,
let $x_0 \in M$ denote an initial point and $v_0 \in T_{x_0}M \cong\bR^n$ a deviation vector.
These evolve under the system
\bsplit{OneJet}
	x_{t+1} &= f(x_t) ,\\
	v_{t+1} &= Df(x_t)v_t ,
\esplit
for $t \in \bN$.
The Lyapunov exponent for $(x_0,v_0) \in TM$ is then the growth rate of the norm of $v_t$:
\bsplit{LyapDef}
	\mu_T(x_0,x_0) =  \frac{1}{T} \ln\left(\frac{ \|v_T\|}{\| v_0\|}\right) , \\
	\mu(x_0,v_0) = \limsup_{T \to \infty} \mu_T(x_0,v_0)
\esplit
if this limit exists.
Equivalently, the time $T$ exponent can also be written as the time average of the exponential stretching factors
\bsplit{LyapAve}
	\mu_T(x_0,v_0) & 
	                =\frac{1}{T} \sum_{t=0}^{T-1} s_t ,\\
	s_t &\equiv \ln \left( \frac{\|v_{t+1}\|} {\|v_{t}\|} \right) , 
\esplit
since the sum is telescoping.
Note that \Eq{LyapAve} is the time average of the \textit{stretching} function $S: TM \to \bR$ on the tangent
bundle, defined by $s_t = S(x_t,v_t)$, so that
\beq{Stretch}
	S(x,v) \equiv \ln \left(\frac{\|Df(x) v\|}{\|v\|}\right) .
\eeq
Convergence of $\mu_T$ as $T \to \infty$ almost everywhere
with respect to an invariant measure was proven in
Oseledec's multiplicative ergodic theorem under certain restrictions
\cite{Oseledec68, Ruelle79, Raghunathan79, Kuznetsov12}.

A similar process can be used to compute the spectrum of exponents. A standard technique is 
repeated application of Gram-Schmidt orthogonalization \cite{Benettin80c, Dieci95, Alligood97, Udwadia01, Dieci02}.
Given an initial orthonormal basis $Q_0 = (q_0^{(1)},q_0^{(2)}, \ldots q_0^{(n)})$, one iterates and then orthogonalizes:
\[
	\begin{array}{l}
	p^{(j)} = Df(x_t) q^{(j)}_{t}  \\
	z^{(j)} = p^{(j)} - \sum_{k=1}^{j-1} \frac{\langle p^{(j)},z^{(k)}\rangle}{\|z^{(k)}\|^2} z^{(k)}
	\end{array}, 
	   \quad \, j = 1,\ldots n .
\]
Normalization of the orthogonal basis $Z = (z^{(1)},\ldots,z^{(n)})$ then gives the scaling factors
and new orthonormal basis
\[
	r^{(j)}_{t+1} = \| z^{(j)}\| , \quad
	q^{(j)}_{t+1} =  \frac{z^{(j)}}{r^{(j)}_{t+1}} .
\]
Iterating this process along an orbit $\{x_t\}$ gives a sequence $Q_t$ of orthogonal matrices
and growth factors $r^{(j)}_{t}$. The spectrum of Lyapunov exponents then becomes
\begin{eqnarray}\label{eq:LyapSpec}
	\mu^{(j)}(x_0) &=& \limsup_{T \to \infty} \mu_T^{(j)}(x_0) , 	\mbox{ where }\\
 \mu^{(j)}_T(x_0) &=& \frac{1}{T}\sum_{t=0}^{T-1} \ln(r^{(j)}_{t+1}), \nonumber
\end{eqnarray}
when the limit exists.
We define the stretching for the $j^{th}$ exponent by 
\begin{eqnarray}\label{eq:StretchFunction}
	R^{(j)}(x_t,v_t) &=& \ln (r_t^{(j)}), \, \mbox{ and let }\, \\
	 R &=& (R^{(1)}, \dots, R^{(n)}) \, . \nonumber  
\end{eqnarray}

There has been a large amount of research on computing Lyapunov exponents, and more generally on methods
for distinguishing chaotic orbits. Accurate computation of $\mu$ is difficult---the
convergence of the limit \Eq{LyapDef} is often no faster than $\ln(T)/T$ \cite{Cincotta16}.
There have been many attempts to accelerate this convergence \cite{Christiansen97b,Geist90,Rangarajan98, Sandor04}; 
however, such an acceleration seems to be difficult.
Indeed, it often is difficult to even determine if $\mu \neq 0$; for example, orbits that are very close
to regular regions in a Hamiltonian system can be chaotic
but have arbitrarily small maximal Lyapunov exponents \cite{Bountis12}.
Nevertheless, specialized methods may help in some cases, for example, 
dynamics conjugate to an incommensurate rotation \cite{Haro16, Szezech05, Fiedler24}
or random matrix products of shears \cite{Sturman19}. 

Our goal in this paper is to investigate if the weighted Birkhoff average (WBA) \cite{Das17, Sander20, Meiss21}
can help to \textit{accurately} compute Lyapunov exponents.  
As we will see, for smooth maps and nonchaotic orbits, $C^\infty$-smooth, weighted averages can compute
(non-positive) Lyapunov exponents with super-polynomial convergence, meaning with convergence faster than $1/T^k$
for all $k \in \bN$.
On the other hand, just as for functions on phase space \cite{Sander20}, we will see that a weighted average
usually does not improve the rate of convergence of the $\mu^{(j)}_T$ when the orbit is chaotic.

Instead of accurately computing $\mu$, many methods, 
such as frequency analysis, fast Lyapunov indicators, 0-1 Test, SALI, GALI, MEGNO, REM, RLI, etc., have
been developed with a weaker goal: that of distinguishing chaotic from regular motion, see e.g., 
\cite{Laskar93a, Voglis99, Gottwald09, Skokos10, Maffione11, Skokos16, Cincotta16, Moges20}.
Many of these methods have been compared in \cite{Bazzani23}.
However as we showed in \cite{Sander20, Meiss21}, the WBA for a function on phase space 
can efficiently distinguish between regular and chaotic orbits by its convergence rates.
Consequently, if this is the only goal, it would be more efficient to 
compute the WBA for a function since in this case only iterates of the map $f$ and not those 
of its derivatives are needed. In \cite{Sander20} we compared the convergence of the WBA to
that of conventional Lyapunov exponents as well as to the 0-1 test \cite{Gottwald09}.


The paper proceeds as follows. Section \ref{sec:WBA} recalls the weighted Birkhoff average.
In \Sec{MEGNO} we show that the ``mean exponential growth of nearby orbits'' (MEGNO) method
\cite{Cincotta16} can be reformulated as a weighted average method,
but not one that is $C^\infty$ smooth at the endpoints.
In \Sec{results}, we compare five weight functions for estimating the time average in \Eq{LyapSpec}.
Several example maps that we think of as ``typical" are discussed in \Sec{typical}. 
Finally, in \Sec{outliers} we discuss some outliers;
maps which have unexpected speed-up or slow-down of convergence.
These examples include maps with shear, that are noninvertible, and those with fixed Jacobian. 
Our conclusions appear in \Sec{conclusions}.

\section{Weighted average methods}\label{sec:WBA}

In this section, we review weighted average methods. Since \Eq{LyapAve} is a dynamical time average,
its convergence is related to that implied by Birkhoff's ergodic theorem,
which states that time averages equal space averages for $L^1(M,\bR)$ functions on an ergodic invariant set,
see e.g., \cite{Billingsley65}. Unfortunately, the convergence of such averages is typically slow, i.e., no
faster than $1/T$ \cite{Kachurovskii96}.
A technique for accelerating converge of time averages is
the method of the weighted Birkhoff average developed by \cite{Das17}.
In this work, an average like that in \Eq{LyapAve} for the stretching function \Eq{StretchFunction} is replaced by
\beq{WBLyap}
	\WB_T(R)(x_0,v_0) = \sum_{t=0}^{T-1} w_{T}(t) R(x_t,v_t) .
\eeq

Here $w_T: [0,T] \to \bR$ is a normalized weight, which we write in the form
\beq{Normalized}
	w_{T}(t) = \frac{1}{N_T} g\left(\tfrac{t}{T}\right), \quad
	N _T = \sum_{t=0}^{T-1} g\left(\tfrac{t}{T}\right),
\eeq
for an unnormalized weight function $g: [0,1] \to \bR^+$.

In general the acceleration of the convergence of $\WB_T$ for functions on phase space
relies on the fact that $g(\tau )$ is a bump function: 
it vanishes at $0$ and $1$ and is smooth on the closed interval $[0,1]$.
In particular it has been proven that when the orbit lies on an invariant torus on which the dynamics is conjugate
to a rigid rotation with incommensurate frequency, and the map and conjugacy are sufficiently
smooth, then such a bump function improves the convergence of the Birkhoff average of
a sufficiently smooth function on phase space \cite{Das17, Duignan23, Tong24}. In the best
cases, the convergence is super-polynomial (faster than $1/T^k$ for any $k\in \bZ$) or even exponential.

Our goal in the current paper is to investigate when this improvement extends to computations of the Lyapunov spectrum \Eq{LyapSpec}.

\subsection{Bump Functions}\label{sec:Bump}
 
In this paper we will test several weight functions $g$ \Eq{Normalized} to see how they influence the convergence 
of the time average \Eq{WBLyap} for the Lyapunov spectrum.

The standard choice for $g$ is the $C^{\infty}$ bump function of \cite{Das17}
\footnote{Other possibilities with varying ``widths''  were explored in \cite{Duignan23}.
For time averages of functions on phase space, \Eq{Bump} was found to have the best convergence properties.}
\beq{Bump}
	g_{\rm wba}(\tau)  =  \left\{\begin{array}{ll} e^{-\left({\tau(1-\tau)}\right)^{-1}} & \tau  \in (0,1) 
					\\  0 & \tau  = 0,1 \end{array} \right. .
\eeq
We also consider the skewed bump function
\beq{Skew}
	g_{\rm skew}(\tau) =  \left\{\begin{array}{ll} e^{-\left({\tau^2 (1 - \tau^2)}\right)^{-1}} & \tau  \in (0,1) 
					\\  0 & \tau  = 0,1 \end{array} \right. .
\eeq
This is still $C^\infty$, but is no longer symmetric about $\tau = \tfrac12$---it has a maximum
at $\tau = \tfrac{1}{\sqrt{2}}$.
As a third example, we will use the function
\beq{Left}
	g_{\rm left}(\tau) =  \left\{\begin{array}{ll} e^{-\left(\tau(1-\tfrac12 \tau)\right)^{-1}} & \tau  \in (0,1) 
					\\  0 & \tau  = 0,1 \end{array} \right. , \quad
\eeq
which is $C^{\infty}$ and monotone increasing on $[0,1)$ but has a discontinuity at $t=1$.
This will test whether the computations of $\mu_T$ are more sensitive to initial transient behavior in the
stretching of the vector $v_t$. 
These functions are shown in \Fig{WeightFunctions}. 

\InsertFig{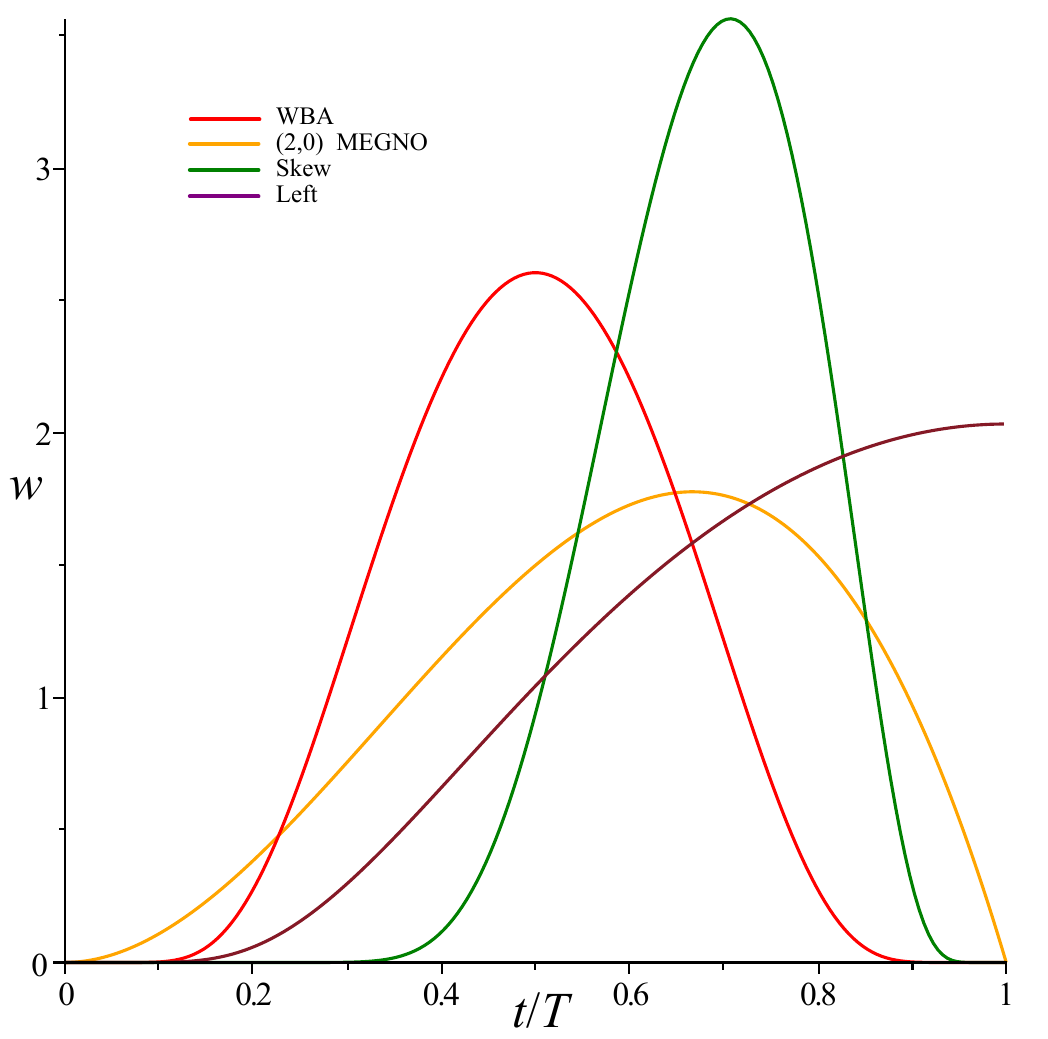}{Normalized exponential bump \Eq{Bump}, $(2,0)$
MEGNO \Eq{Weight20}, ``skew'' \Eq{Skew} and ``left'' \Eq{Left} weight functions.
The weights are normalized to have unit area, 
instead of using the sum as in \Eq{Normalized}.}{WeightFunctions}{3in}

\subsection{MEGNO}\label{sec:MEGNO}

In \cite{Cincotta00}, a modification of the average \Eq{LyapAve} was introduced to give a chaos indicator that
they entitle the ``mean exponential growth rate for nearby orbits'' (MEGNO). 
As reviewed in \cite{Cincotta16}, a generalized MEGNO can be formulated, labeled by a pair integers $(m,n)$.
It is obtained by first computing a weighted average of the stretching factor \Eq{Stretch}
for the first $t$ iterates:~\footnote
{To be consistent with the definition of the stretch \Eq{Stretch}, and the concept of a bump function,
we shift the indices by one step from \cite[Eqs.~4.38-39]{Cincotta16}.}
\[
	Y_{m,n}(t) = (m+1) t^n \sum_{j=0}^{t} j^m S(x_j,v_j) .
\]
The $(m,n)$ MEGNO is obtained as an additional time average of this quantity:
\[
	\bar{Y}_{(m,n)}(T) = \frac{1}{T^{m+n+1}} \sum_{t=0}^{T-1} Y_{m,n}(t).
\]
Note that we can reorder this double sum to obtain an expression that
closely resembles the weighted average \Eq{WBLyap}:
\[
	\bar{Y}_{(m,n)}(T) =  \sum_{t=0}^{T-1} W^{meg}_T(t) S(x_t,v_t) ,
\]
where the MEGNO ``weight" is effectively
\beq{MEGw}
	W^{meg}_T(t) =  \frac{m+1}{T^{m+n+1}} t^m \sum_{j=t}^{T-1} j^n \;.
\eeq
Though $\bar{Y}_{(m,n)}$ now has a form similar to \Eq{WBLyap}, the weight function is \textbf{not} normalized:
MEGNO does not attempt to compute an accurate value for $\mu_T$.
We propose that a normalized weight function would be more appropriate, so we rescale \Eq{MEGw} to define
\beq{WeightM}
	w_{T}^{meg}(t) = \frac{1}{N_T} \left(\frac{t}{T}\right)^m  \frac{1}{T} \sum_{j=t}^{T-1} \left(\frac{j}{T}\right)^n , \\
\eeq
where $N_T$ is the normalization constant as in \Eq{Normalized}. Then the MEGNO-weighted average
for the Lyapunov exponent is defined using this weight in \Eq{WBLyap}.

According to \cite{Cincotta16}, the most useful cases of MEGNO correspond to $(m,n) = (1,-1)$ and $(2,0)$.
We will compare our results only with the second case since it was shown to converge more rapidly.
When $n = 0$ the sum in \Eq{WeightM} is trivial. Normalizing then gives
\beq{MEGwgt}
	w^{meg}_T(t) = 12 \frac{ t^2(T-t)}{ T^2 (T^2-1)} .
\eeq
Note that this function vanishes at $t=0$ and $T$: it is a bump function like those in \Sec{Bump}.
To compare more directly with these, we rescale time and set the interval to $[0,1]$ to obtain
\beq{Weight20}
	g_{meg}(\tau) = \tau^2(1-\tau).
\eeq
and then $w_T^{meg}$ is given by the normalization \Eq{Normalized}.
This weight is $C^1$ at the endpoint $\tau=0$, but only $C^0$ at $\tau=1$.
This function is the orange curve in \Fig{WeightFunctions}.


\section{Numerical results}\label{sec:results}

In this section, we compare the performance of the averages defined in \Sec{WBA} by computing the 
Lyapunov spectrum for several example maps with regular and chaotic orbits. 
The first case, in \Sec{typical}, exemplifies what we believe is the ``typical'' behavior.
We then describe in \Sec{outliers} cases where the behavior is atypical due to special properties of the dynamics. 

We first compute Lyapunov exponents using the Gram-Schmidt method and standard, 
unweighted average \Eq{LyapSpec}. Then we compute the exponents using the
four weighting functions: exponential \Eq{Bump}, skew \Eq{Skew}, left \Eq{Left}, and  $(2,0)$-MEGNO \Eq{Weight20}.
In each case our goal is to understand how the averages converge as $T \to \infty$. 
The errors at time $T$ could be computed if we knew the theoretical values of the exponents,
$\mu^{(j)}$; however, in most cases these are not known. 

Instead, we estimate the exponents by
using the values $\mu_{T^*}^{(j)}$ at a fixed, large  $T^*$ to give an estimate of the ``true'' answer.
In order to avoid bias, rather than choosing a fixed ``truth'', each method produces its own estimate. 
We will say that a Lyapunov exponent \textit{converges as $T^{-k}$} if
\[
	|\WB_T(R) - \mu_{T^*}| \sim T^{-k} \quad \mbox{ for } \quad  1 \ll T \ll T^*,
\]
i.e., if a log-log plot of the error has slope $-k$ over some interval.

\subsection{Typical Convergence}\label{sec:typical} 

Recall that the weighted Birkhoff average $\WB_T(h)$ for a function $h \in C^\infty(M,\bR)$ on phase space 
converges slowly when an orbit is chaotic but for ``regular" orbits
(those smoothly conjugate to incommensurate rotations) it converges
at a rate rate determined by the smoothness of the weight function---for a $C^\infty$ weight, such as \Eq{Bump},
this can be super-polynomial \cite{Das17,Sander20,Meiss21, Tong24}. 
By contrast, the standard unweighted average nominally convergences at best as $T^{-1}$ \cite{Kachurovskii96}.

Here we similarly observe that the Lyapunov spectrum also converges slowly whenever the orbit
is chaotic, regardless of the method used. But for a regular orbit, the weighted averages do typically enhance
the convergence of the Lyapunov spectrum.

As a first example, consider the three-dimensional ``discrete Lorenz map'' \cite{Gonchenko23}
\bsplit{Lorenz}
x' &= y \\
y' &= -0.85 x + \nu_2 y  + y z \\
z' &=  0.95 z - y^2	\;.
\esplit
Here we fix two of the parameters ($\nu_1 = -0.85$ and $\nu_3 = 0.95$, in the notation of \cite{Gonchenko23})
and  allow only the parameter $\nu_2$ to vary.
For \Eq{Lorenz}, the determinant of the Jacobian is independent of the point: $\det(Df) = \nu_1 \nu_3 = -.8075$.
This implies that the sum of the exponents should be
\[
	d^* = \ln(|\det(Df)|) \approx -0.2138122238853254.
\]
However, in the computations below, we do not use this since we want to test convergence of the individual exponents \Eq{LyapSpec}.

Figure \ref{fig:LyapLorenzPar} shows the three Lyapunov exponents for this map as a function of $\nu_2$, computed 
using the standard WBA weight \Eq{Bump} with the iterative Gram-Schmidt method \Eq{LyapSpec}.
The orbit is arbitrarily chosen to start at $(0,-0.01,0.0001)$, discarding
the first $4000$ iterates to remove transients. The exponents are computed
using \Eq{WBLyap} for the next $T = 2(10)^4$ iterates.
In all cases,  $|\mu^{(1)} + \mu^{(2)} + \mu^{(3)} - d^*| < 10^{-14}$, 
consistent with the constant Jacobian determinant.
This excellent convergence follows from the fact that, up to floating point error, 
$\ln(r_t^{(1)})+ \ln(r_t^{(2)})+\ln(r_t^{(3)}) = d^*$ for each
$t$, so the sum of the averages is the average of the sum, with or without a weight function.

\InsertFig{LyapLorenzPar}{The three Lyapunov exponents for the discrete Lorenz map for $1000$
values of $\nu_2  \in [ 1.83, 1.95]$.
These were computed using WBA weight \Eq{Bump} with $T= 2(10)^4$.
The dashed lines mark the values of $\nu_2$ used in \Fig{DiscreteLorenz}.}{LyapLorenzPar}{3in}

As shown in \cite{Gonchenko23}, the fixed point $(x,y,z) = (0,0,0)$ of the map \Eq{Lorenz} is
stable up to $\nu_2 = 1.85$ where it undergoes
a pitchfork bifurcation---at this point the largest exponent (blue in the figure) hits zero.
The newly created pair of fixed points lose stability at
$\nu_2 \approx 1.8645$ in Neimark-Sacker bifurcations. The resulting pair of attracting circles
or periodic orbits have basins of attraction limited by the stable and unstable manifolds of the origin,
and there can be additional attractors.
At $\nu_2 = 1.87$ there is a chaotic, Lorenz-like attractor, as shown in \Fig{DiscreteLorenz}(a).
This attractor has a single positive Lyapunov exponent and a tangential exponent of zero.
Using the exponential weight \Eq{Bump} with $T^*=2(10)^6$ gives the exponents
\[
	\mu \simeq (0.0039858,0.0000000,-0.2177981). 
\]
As $\nu_2$ varies, the Lorenz-like attractor can collapse onto an attracting circle;
for example this occurs at $\nu_2 = 1.8785$, see \Fig{DiscreteLorenz}(c).
For this attracting circle the maximal Lyapunov exponent is zero, and
again using the exponential weight \Eq{Bump}, we find (to higher accuracy)
\[	\mu \simeq \left( 
\begin{array}{c}
0.000000000000000 \\ 
-0.000160991051261 \\ 
-0.213651232834031 
\end{array} 
\right) ,
\]
for the same $T^*$.
These cases are marked with vertical dashed lines in \Fig{LyapLorenzPar}.

The convergence of the largest exponent using the five weight functions of \Sec{WBA} is shown in 
\Fig{DiscreteLorenz}, panels (b) and (d), for $\nu_2 = 1.87$ and $1.8785$, respectively.
These plots show the errors for $100$ logarithmically spaced values of $T \in [900,1.5(10)^6]$.
In each case the first $4000$ iterates of the initial condition are discarded to remove transients.
The error is estimated by comparing to $\mu^{(1)}_{T^*}$ with $T^* = 2(10)^6$. 

As seen in \Fig{DiscreteLorenz}(b), all of the methods perform poorly for the chaotic
attractor at $\nu_2 = 1.87$: the convergence is at best like $T^{-1}$.
Consistent with this, one might believe the results to 5 or 6 digits at $T= 10^6$.
The convergence of the second and third Lyapunov exponents (not shown) is 
essentially indistinguishable from that shown in panel (b). 

For the invariant circle at $\nu_2 = 1.8785$, the WBA and skew weights, which are $C^\infty$,
far outperform the other methods. The best
convergence is for the skew weight; it essentially reaches machine precision by $T= 3(10)^5$.
The MEGNO weight \Eq{Weight20} also gives increased convergence at a rate nearing $T^{-2}$. The left and constant 
(regular) weights still converge as $T^{-1}$: there is no improvement since these weight functions are not continuous. 

\InsertFigFour{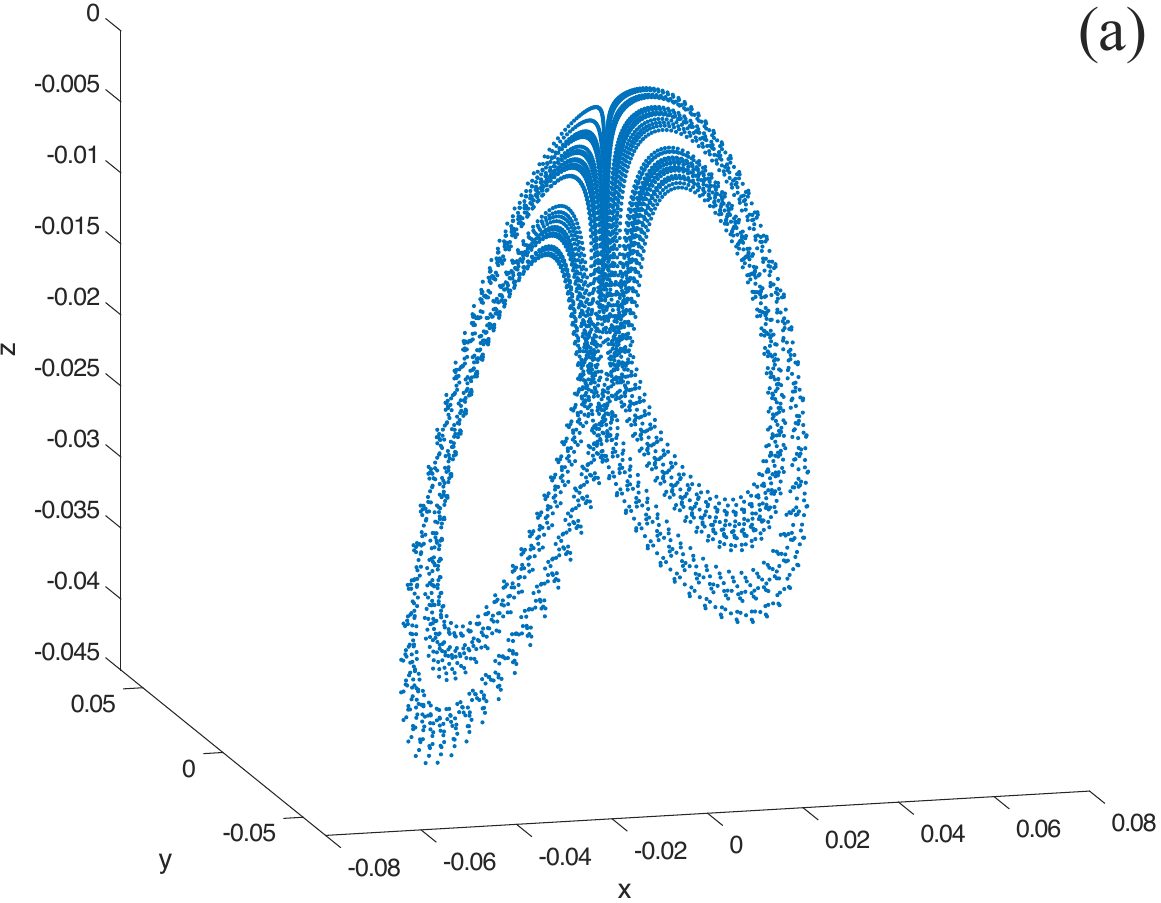}{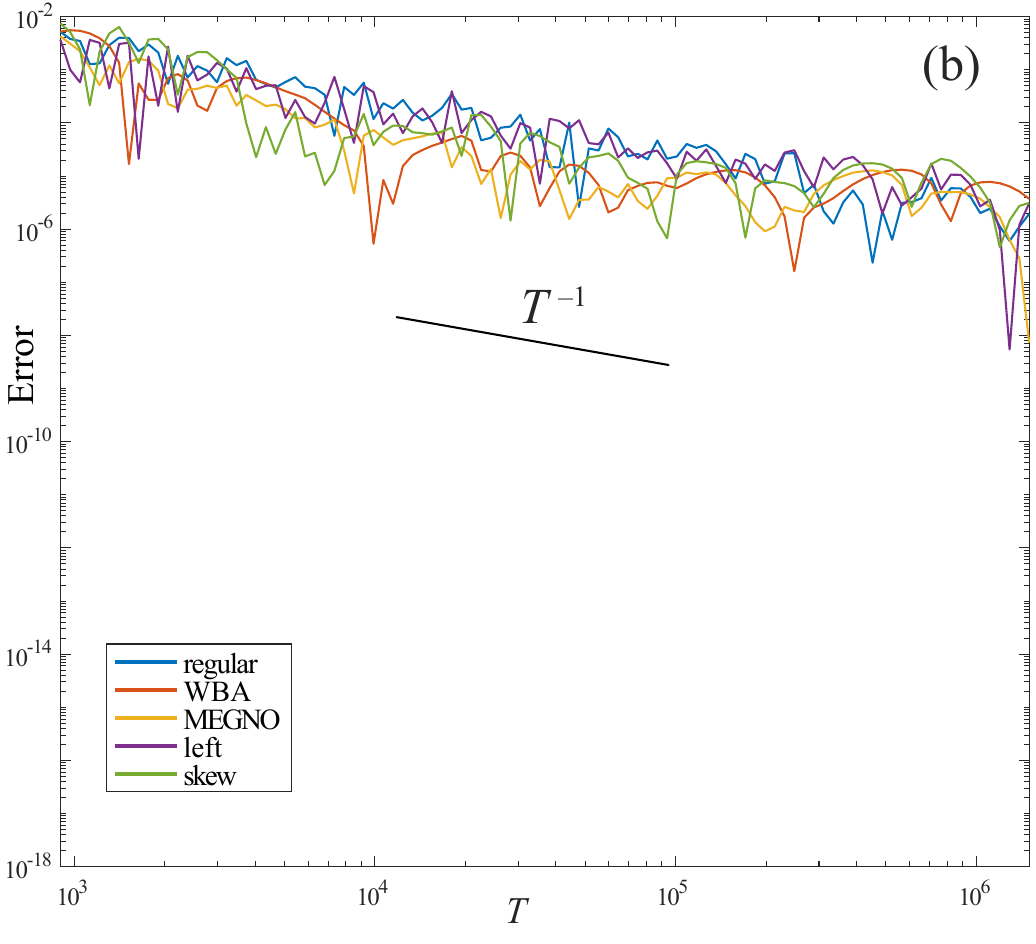}{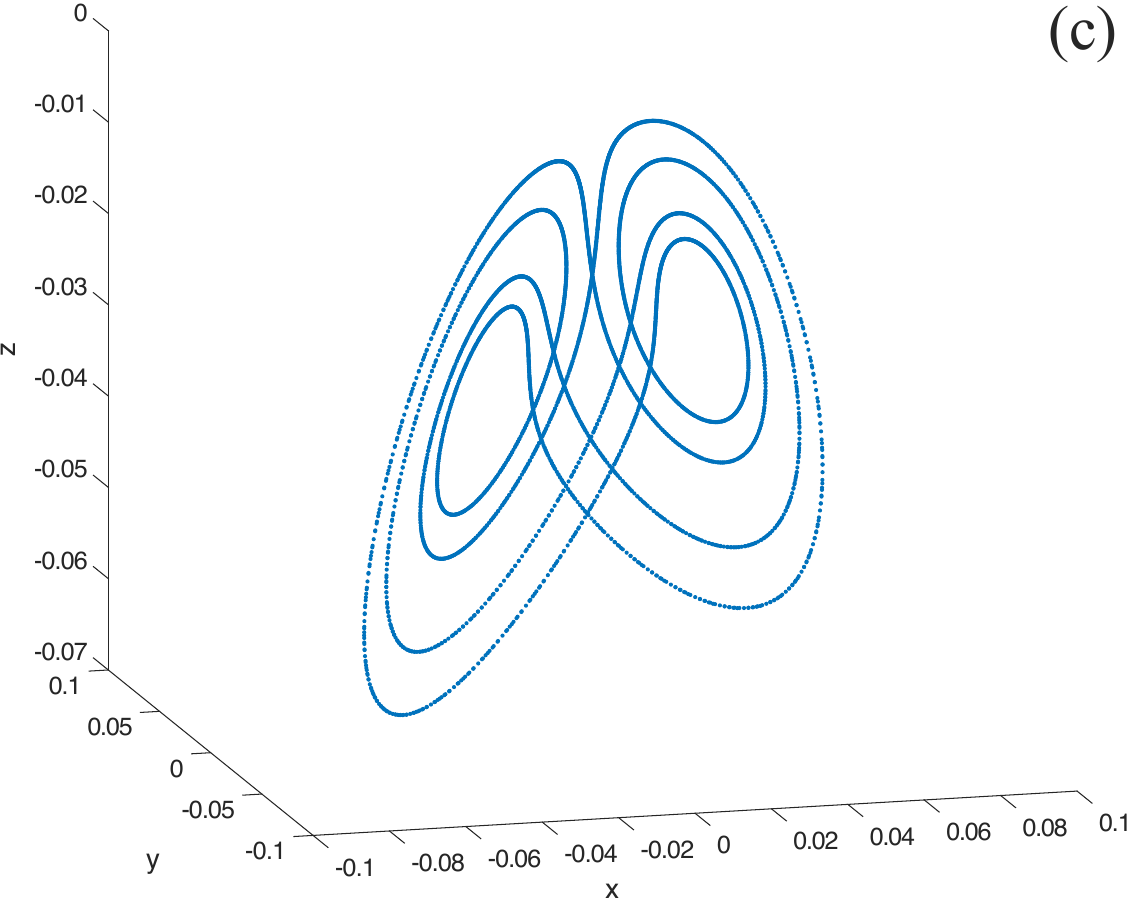}{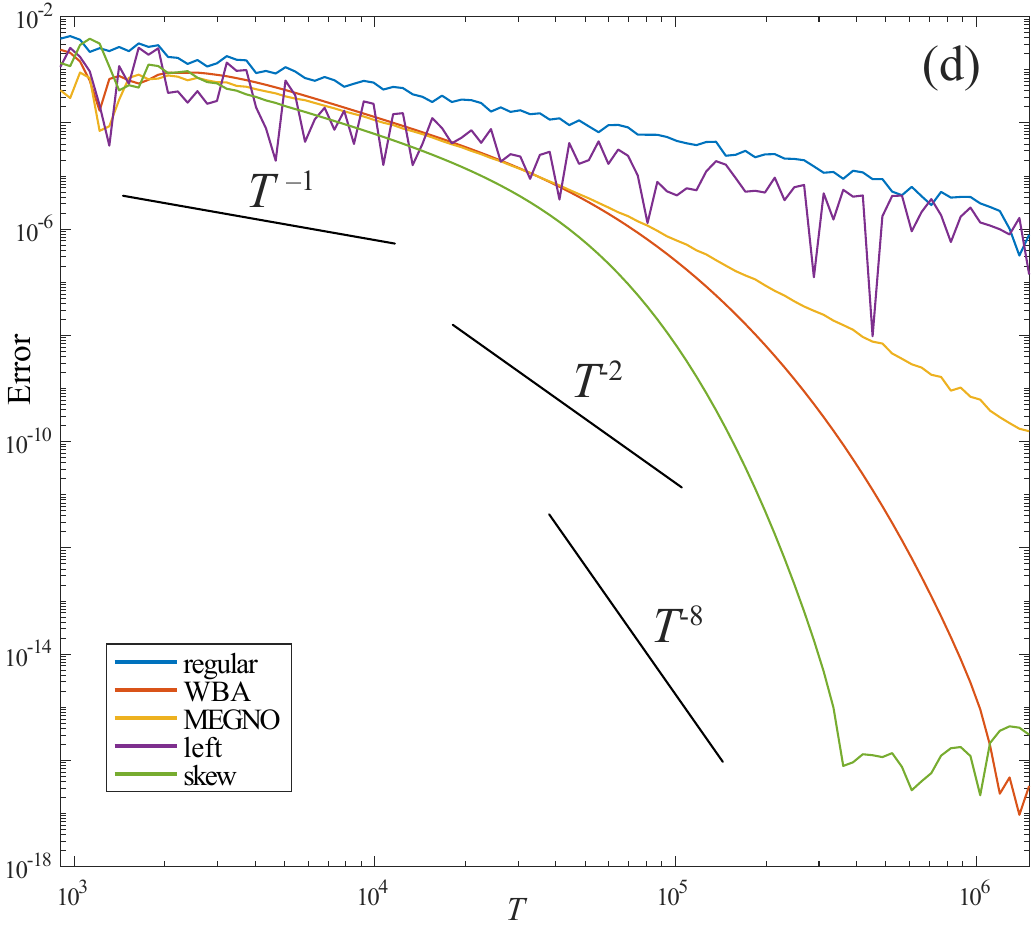}{Two attractors for the discrete Lorenz map \Eq{Lorenz}
and corresponding errors for the largest Lyapunov exponent.
Panels (a) and (b): Chaotic attractor with $\nu_2 = 1.87$.
The weighted averages show no improvement over the ``regular'' method.
Panels (c) and (d): Invariant circle at $\nu_2 = 1.8785$.
The smoothly weighted methods converge much more quickly.
The curves in panels (b) and (d) correspond to the unweighted average (regular), exponential bump (WBA) \Eq{Bump},
$(2,0)$-MEGNO \Eq{Weight20}, left \Eq{Left}, and skew \Eq{Skew} weights as labeled.
}{DiscreteLorenz}{2.6in}

The behavior seen in \Fig{DiscreteLorenz} appears to be typical: we have seen similar performance for
convergence of the Lyapunov spectrum in many other simulations for chaotic and nonchaotic orbits 
for various maps including the discrete Lorenz map for other parameter values, the classic H{\'e}non map, the 
Derived from Anosov (DA) map \cite{Coudene06}, and the 2D torus map studied in \cite[Section 3.7.3]{Das17}.
Similar behavior is also seen for Poincar\'e maps of flows, such for a periodically forced, double-well Duffing oscillator \cite{Kanamaru08}.

\subsection{Outliers}\label{sec:outliers}

We now describe several cases where the convergence does not follow the pattern seen in \Sec{typical}.

{\bf Dynamics with Shear:}
Integrable symplectic maps have families of invariant tori whose rotation vectors generically vary across tori: they have \textit{shear}. This results in the linear growth of the length of vectors transverse to the tori,
and this gives the slow convergence 
\[
	\mu_T \sim \frac{ \ln(T)}{T}
\]
to zero for \Eq{LyapAve}.
It has long been known \cite{Casati80} that shear causes a similar slow convergence
even when the map is not integrable, whenever the orbit lies on an invariant torus \cite{Skokos16}. It appears to be better for the weighted cases, as we describe below and see in \Fig{VDP}.

For example, consider the 3D volume-preserving map for $(x,y,z) \in \bT^2 \times \bR$ \cite{Meiss21}:
\bsplit{VPMap}
	x' &= x + z'  + \tfrac12 (\sqrt{5}-1)  \mod 1\\
	y' &= y + 2 (z')^2 + 0.4 \mod 1\\
	z' &= z - 0.02 \left( \sin(2 \pi x)  + \sin (2 \pi y) + \sin(2 \pi (x-y) \right) \;.
\esplit
The initial point $(x,y,z)= (0,0,-0.05)$ appears to lie on 
an invariant two-torus that is a graph over $(x,y)$ on which the dynamics
has the incommensurate rotation vector $\omega \simeq (0.544519,0.411571)$ \cite{Meiss21}.
All three of the Lyapunov exponents for this orbit are zero: the two tangential exponents
vanish because the invariant set is a two-torus, and the transverse exponent is then zero because
of volume-preservation. Nevertheless, since the rotation vector varies with $z$, the map has shear and
the length of a vector transverse to the torus will grow linearly.
This should result in slow convergence of the exponents.

The convergence of a Birkhoff average of the function $h = \cos (2 \pi x)$ and of the largest
Lyapunov exponent are shown in \Fig{VDP} for the five weight functions of \Sec{WBA}.
The figure shows the averages for $100$ values of $T \in [800, 1.5(10)^7]$.
Since the map is volume preserving, no transient removal is needed.
For the $C^\infty$ weight functions the convergence rate of the Birkhoff average is excellent,
as we would expect for a quasiperiodic orbit \cite{Meiss21}. 
The errors for the Lyapunov exponent in panel (b) were computed by comparing to $\mu^{(1)} = 0$.
Note that this convergence is very slow---even though the orbit is nonchaotic.
The other two Lyapunov exponents (not shown) also have the same convergence rates.

We observe similar behavior for other parameters and orbits of the map \Eq{VPMap}
as well as for nonchaotic orbits of the 2D Chirikov standard map.

\InsertFigTwo{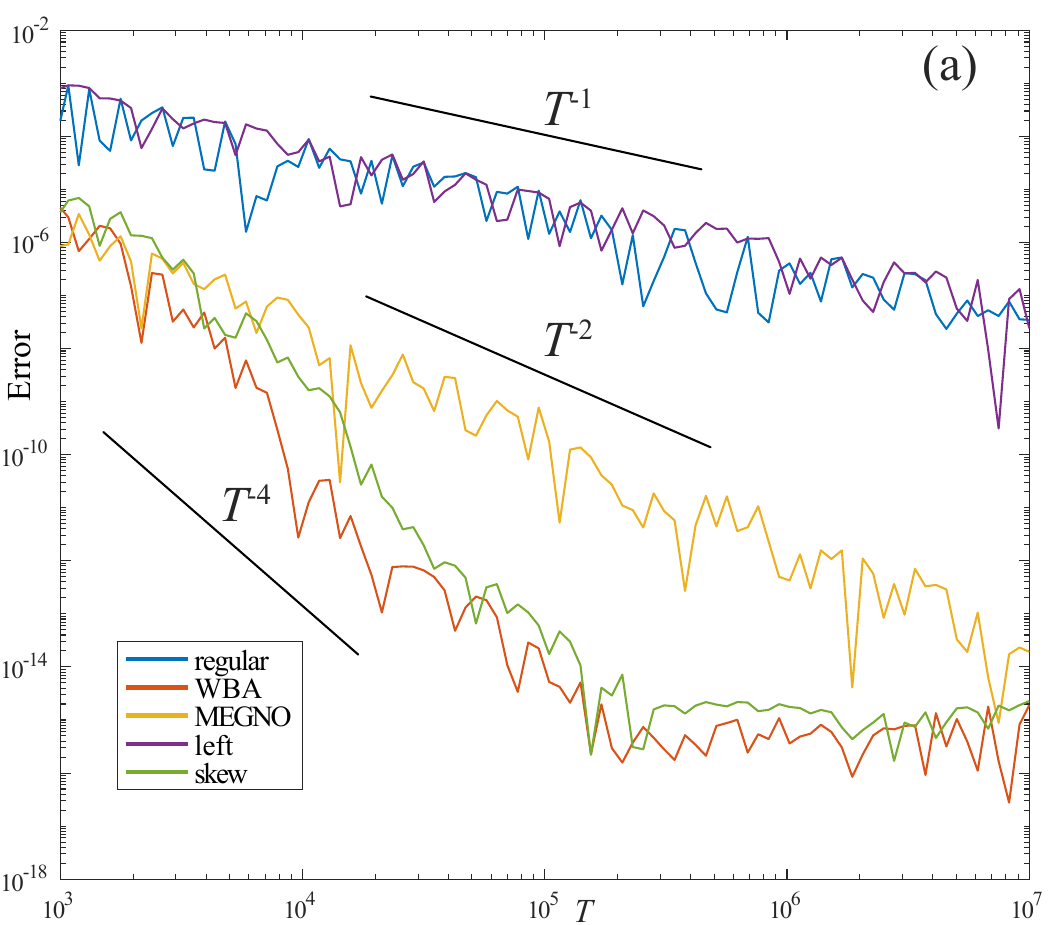}{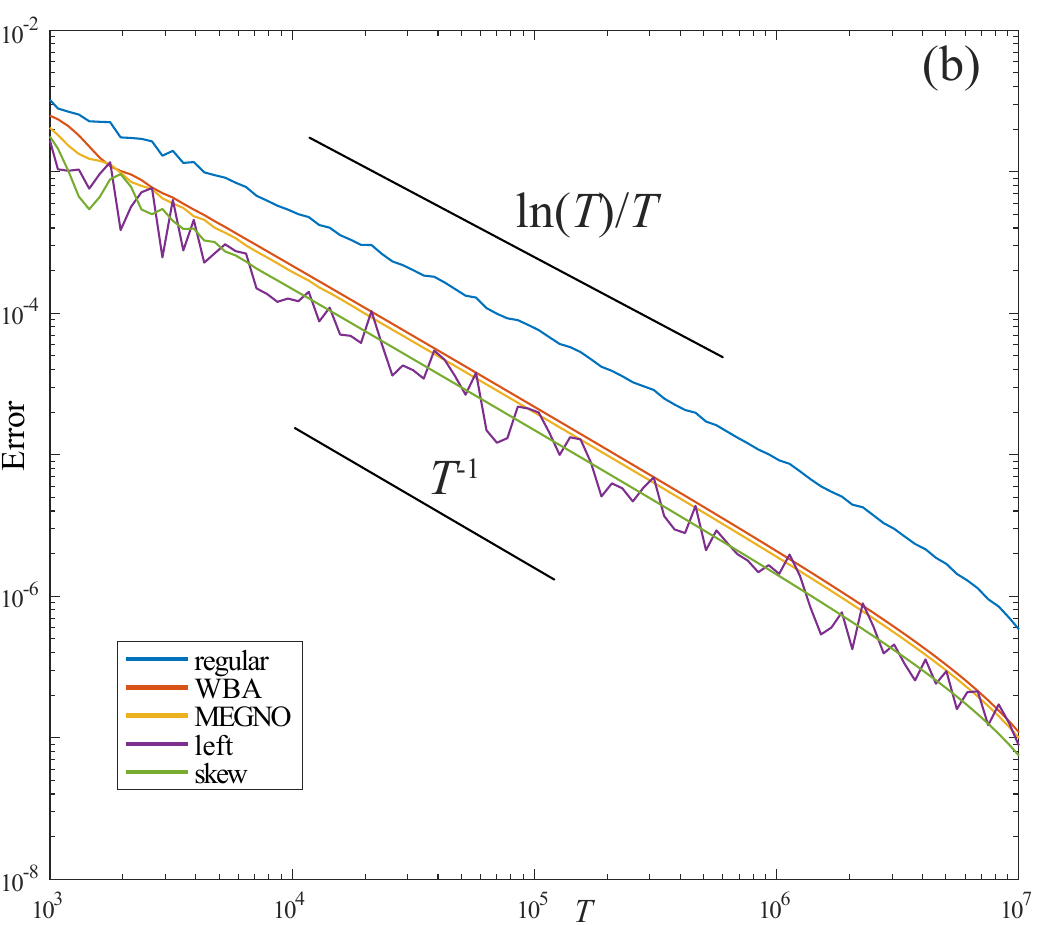}{Convergence of weighted averages for an invariant torus of the map \Eq{VPMap}.
Panel (a): time averages of $\cos(2 \pi x)$. For the $C^\infty$ weights this average converges to machine precision
by $T \approx 3(10)^5$. Panel (b): convergence of a Lyapunov exponent to zero. This appears to converge as $\ln T/T$ 
for the regular average and as $1/T$ for the weighted averages. 
Note the vertical scales are different in the two panels.}{VDP}{3in}

{\bf Weak Chaos:} 
A dynamical system has {\em weak chaos} when the Lyapunov exponents are not positive,
but it still has sensitive dependence on initial conditions. Such dynamics
can lead to strange nonchaotic attractors (SNA), where the orbit lies on a geometrically
strange (fractal) set, \cite{Ding89a, Glendinning06}. In this case we have previously observed that averages
converge slowly for both Lyapunov exponents and functions on phase space, and
adding a weight function does not improve convergence \cite{Duignan23, Meiss24}.

{\bf Noninvertible maps:}
Noninvertibility makes the computation of Lyapunov exponents more delicate. 
Denote the set of points where the Jacobian of the map is singular by
\[
	J_0 = \{ x: \det(Df(x)) = 0\}.
\]
If an invariant set intersects $J_0$, it may be nonsmooth and even self-intersecting,
see \cite{Sander99, Sander00, Josic04} and references therein.
If an ergodic component intersects $J_0$, even if it is smooth, some of the exponents will be undefined,
since a zero determinant implies that $r_t^{(j)} = 0$ for some $j$, so that $\ln r_t^{(j)}$ is infinite.
Moreover, the average \Eq{LyapSpec} will also be undefined for initial conditions on the dense, countably infinite set of 
preimages of $J_0$. Of course it is still possible for the Lyapunov exponents to exist almost everywhere. 
However, from the numerical standpoint, if an orbit nears the dense set of singular points, then at least one
$r_t^{(j)} \approx 0$, and this will lead to significant floating point errors. 

This phenomenon is shown in \Fig{TinkerBell} for a so-called Tinkerbell map \cite{Alligood97}
\bsplit{TinkerBell}
x' &=  0.33 x - 0.6y + x^2-y^2 \\
y' &=  2 x + 0.5 y + 2 xy \;.
\esplit
In this case $J_0$ is a circle centered at $(-0.2,-0.65)$ with radius $\sqrt{0.125}$ that intersects
the attractor, see \Fig{TinkerBell}(a).
The Lyapunov exponents for this attractor are nonpositive, $\mu \simeq (0,-0.14292)$,
and as seen in \Fig{TinkerBell}(b),
the convergence of $\mu^{(1)}$ for the $C^\infty$ bump functions is excellent:
since $J_0$ intersects the orbit transversally, $r_t^{(1)}$ is never near zero.
However, $r_t^{(2)}$ does get arbitrarily close to zero infinitely often on the orbit.
As seen in \Fig{TinkerBell}(c), this results in slow convergence of the second Lyapunov exponent for all methods.
For this case, the attractor is smooth, and we observe that the convergence of $\WB_T(h)$ using \Eq{Bump} for 
functions on phase space (such as $h = \cos(2 \pi x)$ and the rotation number) is excellent (not shown).
The implication is that for this case, the numerical difficulities are restricted to the smaller Lyapunov exponent.

\InsertFigTwo{tinkerbellOrbit}{tinkerbellLyap}{For the Tinkerbell map, the left panel depicts the attractor,
the set of points with a singular Jacobian ($J_0$), and the first iterate of these points ($J_1$).
The right panels show the errors for $\mu_T^{(1)}$ and  $\mu_T^{(2)}$ as a function of $T$
(note the difference in vertical scales).
The second exponent converges slowly for all methods.}{TinkerBell}{3in}

{\bf Constant Jacobian:} When the map $f$ has a constant, hyperbolic Jacobian, the convergence rates of $\WB_T(S)$, \Eq{WBLyap} are enhanced for smooth weight functions. Indeed, when $Df = A$ is constant, the method given in
\Eq{LyapSpec} is equivalent to a computing the singular values $\sigma_j$ of $A$
via normalized simultaneous power iteration. 
As long as  $0 \le \sigma_{j+1} < \sigma_j  <  \dots < \sigma_1$ (the strict inequality is the generic case),
we have $r_t^{(j)} \to \sigma_j$ with error $\cO(|\sigma_{j+1}/\sigma_j|^t)$ \cite{Sauer18}.
As a result, the errors in the weighted average \Eq{LyapSpec},
occur only for the initial transients.
The implication is that the left \Eq{Left} and skew \Eq{Skew} weights perform
better than the others, since they suppress the initial portion of the average.
In contrast, when the orbit is chaotic, $\WB_T(h)$ for functions $h$ converges slowly for any weight function.

We have verified this improved convergence of $\mu^{(j)}_T$ and slow convergence of
other averages for examples including Arnold's cat map
and the skinny baker map \cite{Alligood97}, both of which are uniformly hyperbolic.

\section{Conclusions}\label{sec:conclusions}

The computation of the Lyapunov exponent for orbits of a dynamical system can be formulated as a
time average of the stretching function $S(x_t,v_t)$ along a trajectory.
As noted in \cite{Das17}, these averages can be computed using bump functions, similar to those used
to compute Birkhoff averages of functions on phase space.  An advantage of this smoothing 
is that the results can be super-polynomially convergent when the trajectory is regular.
We extended these ideas to compute the full Lyapunov spectrum \Eq{LyapSpec} using weighted averages \Eq{WBLyap}.

In \Sec{results} we showed that a $C^\infty$ weight function typically gives super-convergence
of the Lyapunov spectrum on nonchaotic orbits, just as it does for functions on phase space.
There are exceptions, however, including invariant sets that are tori with transverse shear;
these are common in the Hamiltonian or symplectic case. Moreover, having nonpositive
Lyapunov spectrum is not sufficient for super-convergence, as the example of weak-chaos shows.
In addition, attractors of non-invertible maps, even if they are non-chaotic, can have slow
convergence of some exponents due to singularities. 
Finally, convergence of exponents can be enhanced by a smooth weight for the simplest of chaotic systems: those that
are uniformly hyperbolic with constant Jacobian.

As we showed in \Sec{WBA}, the MEGNO chaos detection method of \cite{Cincotta00, Cincotta16}
can be reformulated as a weighted time average which gives the maximal Lyapunov exponent.
The weight function in this case, however, is not $C^\infty$ and this results in slower convergence
of the average. Since $C^\infty$ weight functions have essentially the same computational cost as less smooth functions,
there seems to be no reason to use a less smooth weight. 

The results in this paper still leave open the question: is is possible to devise a technique for efficiently and 
accurately computing the Lyapunov spectrum for a typical, chaotic invariant set?

%

\end{document}